\documentclass[prb,twocolumn,showpacs,preprintnumbers]{revtex4}

\usepackage[dvipdfm]{graphicx}
\usepackage{amsmath}
\usepackage{amssymb}

\def\b{\beta}
\def\c{\chi}

\def\e{\epsilon}

\def\s{\sigma}
\def\t{\tau}
\def\w{\omega}
\def\ua{\uparrow}
\def\da{\downarrow}

\def\Vec#1{\mbox{\boldmath $#1$}}
\begin{document}

\title{Itinerant ferromagnetism in the multiorbital Hubbard model
:\\
a dynamical mean-field study}

\author{Shiro Sakai,$^{1*}$ Ryotaro Arita,$^{2}$ and Hideo Aoki$^{1}$}

\affiliation{$^1$Department of Physics, University of Tokyo, Hongo,
Tokyo 113-0033, Japan\\
$^2$Condensed Matter Theory Laboratory, RIKEN, Wako, 
Saitama 351-0198, Japan.}

\date{\today}

\begin{abstract}
  In order to resolve the long-standing issue of how the 
  itinerant ferromagnetism 
  is affected by the lattice structure and Hund's coupling, 
  we have compared various three-dimensional lattice structures 
  in the single- and multiorbital Hubbard models 
  with the dynamical mean-field theory with an improved 
  quantum Monte Carlo algorithm that preserves the spin-SU(2) symmetry.
  The result indicates that {\it both} the lattice structure 
  and the $d$-orbital degeneracy are essential for
  the ferromagnetism in the parameter region representing
  a transition metal.
  Specifically, (a) Hund's coupling, despite the common belief, 
  is important, which is here identified to come from 
  particle-hole scatterings, 
  and (b) the ferromagnetism is a correlation effect (outside the Stoner 
  picture) as indicated from the band-filling dependence.
\end{abstract}
%%%%%%%%%%%%%%%%%
\pacs{71.10.Fd; 75.20.En; 75.40.Mg}
%%%%%%%%%%%%%%%%%
\maketitle

Itinerant ferromagnetism in transition metals and their compounds
has been one of the central issues in condensed-matter physics.
Despite a long history of investigations, dating back to
e.g., the Stoner\cite{s46} or Slater-Pauling\cite{s37} theories,
a full consensus on the mechanism has not been achieved yet.
While the single-orbital Hubbard model was introduced as
the simplest model to capture the ferromagnetism from 
electron-electron interactions, 
it has become increasingly clear that the model 
on usual lattices does not show ferromagnetism
for realistic values of the Coulomb repulsion $U$.
Intensive studies have then ensued to incorporate 
the ingredients other than the Hubbard $U$. 
Two factors are now considered to be significant: 
lattice structure and the degeneracy of $d$ orbits.

Early theories due to 
Kanamori\cite{k63} and Gutzwiller\cite{g63} already suggested 
that lattice structure 
(which dictates the shape of the density of states (DOS)),
is crucial for the ferromagnetism.
Specifically, they discussed the itinerant ferromagnetism 
for a face-centered cubic (fcc) lattice as in Ni 
where the DOS is peaked at the band edge, 
and argued that the peak stabilizes the ferromagnetism.
The problem is revisited in recent studies:
When the orbital degeneracy is ignored, 
Arita {\it et al.}\cite{ao00} showed 
with the fluctuation exchange (FLEX) 
and the two-particle self-consistent approximations that 
a ferromagnetic tendency is strongest for fcc.  
The ferromagnetism on fcc lattices has also been found 
by Ulmke\cite{u98} with the dynamical mean-field theory (DMFT).\cite{mv89}
The importance of lattice structure has also been suggested
by some exact results for flat-band systems.\cite{mt89}

On the other hand, the importance of the $d$-orbital degeneracy 
and the associated Hund exchange coupling 
has long been stressed.\cite{s36} 
While Hund's coupling is an intra-atomic 
interaction favoring aligned electron spins, 
the interaction may cause a long-range ferromagnetic order
through electron transfers.
This was followed by intensive studies, but 
the complexity of a multiorbital model has 
limited the existing studies to some restrictions: 
The two-orbital Hubbard model in one dimension (1D)
has been most intensively studied,
where a ferromagnetic ground state with an antiferro-orbital order is 
expected for the quarter filling ($n=1$; one electron per site), 
for strong repulsion $U$ and Hund's coupling $J$.
Indeed, quantum Monte Carlo (QMC)\cite{gs87}, 
exact diagonalization (ED),\cite{ka94,ks97,h97} 
and density-matrix renormalization-group (DMRG)\cite{sm02} studies
for finite-size chains have confirmed the ferromagnetic
 ground state.
However, these quarter-filled systems are {\it insulating}.
The purpose of the present letter is to look into 
{\it metallic} ferromagnetism in multiorbital systems,
with our eyes set on transition-metal ferromagnets such as Ni.

Several studies on metallic ferromagnetism 
in multiorbital systems exist.\cite{ka94,sm02,fk97,mk98,hv98,footnotebw} 
Numerically exact results for the double-orbital Hubbard model in 1D
have been obtained with ED\cite{ka94,h97} or DMRG\cite{sm02}, 
while those on the infinite-dimensional hypercubic\cite{mk98} 
or Bethe\cite{hv98} lattices have been studied with the DMFT.
These studies have found itinerant ferromagnetism away from the quarter filling
only for very large Hund's couplings $J\gtrsim W/2$ ($W$: bandwidth), 
and no ferromagnetism has been found in a realistic range of $J$
for transition metals.
However, it is in our view still an open question whether Hund's coupling
is essential in real ferromagnets, since these 
calculations do not take account of 
lattice structures of real materials.\cite{footnotebw}

This has motivated us to investigate here the effect of 
lattice structures and of 
multiorbital correlations with the DMFT, comparing a
simple-cubic and fcc lattices in 
the single- and multiorbital cases. 
We shall conclude that 
{\it both the lattice structure and Hund's coupling} are crucial for 
itinerant ferromagnetism in realistic ($\sim$ transition-metal) 
parameter regions.  
Physically, we observe that
(i) particle-hole scatterings, 
neglected in Kanamori's $T$-matrix theory, are 
essential in fact in the presence of Hund's coupling, 
and (ii) 
the present result for the band-filling dependence of the magnetism 
is totally out of Stoner's picture.

\begin{figure}[t]
  \center{\includegraphics[width=6cm, height=4.2cm]{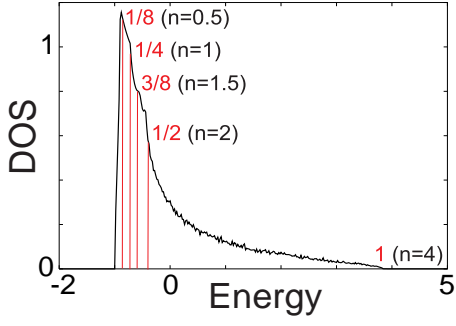}}
  \caption{(Color online) 
           The density of states for noninteracting electrons on 
           the fcc lattice.
           Red figures represent the band filling (per spin), 
           while black ones the corresponding total band filling 
           for a two-orbital system.}
  \label{fig:fccdos}	
\end{figure}

So we start from the multiorbital Hubbard model
on three-dimensional lattices.  
For the fcc lattice the dispersion is
 \begin{eqnarray}\label{eq:disp_fcc}
   \e(\Vec{k})=4t\sum_{i<j}\cos(k_i)\cos(k_j)+2t'\sum_{i=1}^3\cos(2k_i),
 \end{eqnarray}
where $t (t')$ is the nearest-neighbor (second-neighbor) 
hopping, and a cubic Brillouin zone 
($-\pi <k_i\leq \pi$) for two equivalent, 
interpenetrating fcc lattices is adopted.  
We set $t=4t'= \frac{2\sqrt{2}}{3\sqrt{11}}$ 
to fix the effective bandwidth at 
$W\equiv 4(\int \e^2 D(\e)d\e)^{1/2}=4$.
For simplicity we have assumed isotropic 
transfer integrals and ignored hybridizations between orbitals.  
The DOS for noninteracting electrons on fcc 
in Fig.~\ref{fig:fccdos} has a peak at the lower band edge, 
where $E_F$ resides around the peak in Ni in the hole picture.
For comparison we also consider the simple-cubic lattice, 
for which the dispersion is 
$\e(\Vec{k})=2t\sum_{i=1}^3 \cos(k_i)$
with $t=\frac{1}{\sqrt{6}}$ and $W=4$.

The interaction Hamiltonian is
 \begin{eqnarray}\label{eq:Hint}
  &\hat{H}_{\rm int}&\!\!\! \equiv \hat{H}_U + \hat{H}_J, \\
  &\hat{H}_U& \!\!\!\equiv U\sum_{im} n_{im\ua}n_{im\da} \nonumber\\
   &&  \!\!\! +\sum_{i,m<m',\s}\!\![ U'n_{im\s} n_{im'-\s}
                             +(U'-J)n_{im\s} n_{im'\s} ],\nonumber\\
  &\hat{H}_J& \!\!\!\equiv 
   J\!\!\!\sum_{i,m\neq m'} \!\!(
    c_{im\ua}^\dagger c_{im'\da}^\dagger c_{im\da}c_{im'\ua}
   +c_{im\ua}^\dagger c_{im\da}^\dagger c_{im'\da} c_{im'\ua}),\nonumber
 \end{eqnarray}
where $c_{im\s}^\dagger$ creates 
an electron with spin $\s$ in orbital $m$ at lattice site $i$, 
$n_{im\s}\equiv c_{im\s}^\dagger c_{im\s}$, 
$U (U')$ denotes the intra- (inter-)orbital Coulomb interaction, 
and $J$ the Hund-exchange and pair-hopping interactions. 
Here we have decomposed $\hat{H}_{\rm int}$ into 
$\hat{H}_U$ (the density-density interactions) and 
$\hat{H}_J$ (not expressible as such).

We investigate the above model with the DMFT, 
which neglects the $\Vec{k}$-dependence in the self-energy 
and in the vertex but incorporates their temporal dependence and 
the one-electron dispersions.
Since the DMFT becomes exact in the limit of large coordination numbers, 
the approximation is expected to be fair for fcc, 
where a site has twelve nearest neighbors and six second neighbors.
The DMFT impurity problem is solved exploiting the QMC method
developed in our previous paper.\cite{sa06}
The QMC algorithm combines the Trotter decomposition for $\hat{H}_U$
and a series expansion for $\hat{H}_J$
to decouple the two-body interactions with the
auxiliary-field transformations.\cite{h83}
While the algorithm uses a series expansion,
it is virtually nonperturbative, 
since all the nonvanishing orders are incorporated numerically.
An important virtue of the present QMC is that 
it preserves the spin (SU(2)) and orbital ($U=U'+2J$) rotational 
symmetries in $\hat{H}_{\rm int}$, 
which are difficult to preserve in the conventional QMC method.\cite{hv98,hf85}
The present method also enables us to 
address temperatures close to the 
Curie temperature $T_C$, to which the ED or the DMRG cannot access.

We calculate the spin susceptibility,
 \begin{eqnarray}\label{eq:spinsus}
   \c(\Vec{0},0)&\equiv&\sum_{mm'}\c_{mm'}^{zz}(\Vec{0},0),\nonumber\\
   \c_{mm'}^{zz}(\Vec{q},i\nu)&\equiv&
   \int_0^\b d\t  
   \langle {\rm T}_{\t} S_{\Vec{q}m}^z(\t) S_{-\Vec{q}m'}^z(0) \rangle
   e^{i\nu\t},\nonumber\\
   S_{\Vec{q}m}^z&\equiv& \frac{1}{2}\sum_{\Vec{k}ss'}
   c_{\Vec{k}s}^\dagger \s_{ss'}^z c_{\Vec{k}+\Vec{q}s'},\nonumber
 \end{eqnarray}
in the paramagnetic phase ($T>T_C$) in the two-orbital Hubbard model,
through the Bethe-Salpeter equation,
 \begin{eqnarray}\label{eq:bs}
   \hat{\c}^{-1}= \hat{\c}_0^{-1}-\hat{\Gamma},
 \end{eqnarray} 
where $\hat{\c}$ is the susceptibility matrix with respect to 
$m,m',i\w,i\w'$, 
$\hat{\c}_0$ the irreducible lattice Green function 
and $\hat{\Gamma}$ the vertex function.
The DMFT approximates the electron self-energy 
$\Sigma(\Vec{k},i\w)$ and the vertex function 
$\Gamma(\Vec{k},i\w;\Vec{k}',i\w')$ to be 
local, $\Sigma(i\w)$ and $\Gamma(i\w,i\w')$ 
respectively, which are obtained from a QMC calculation 
for the self-consistently-determined impurity model.

\begin{figure}[t]
  \center{\includegraphics[width=7cm, height=10cm]{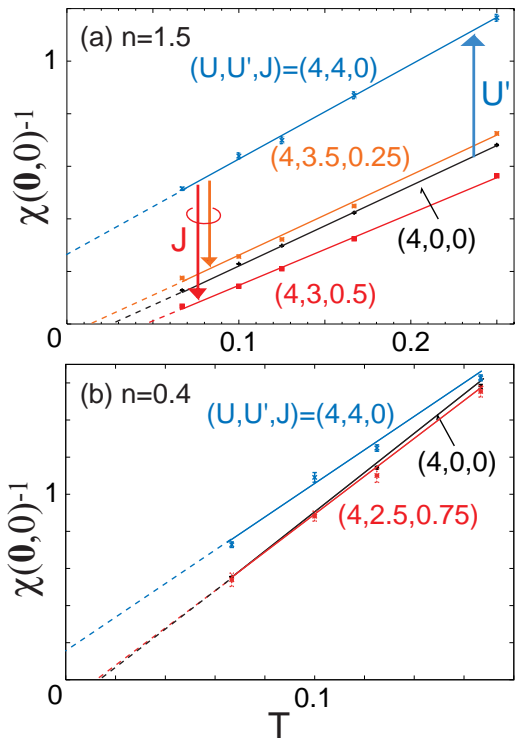}}
  \caption{(Color online) 
           Inverse of the spin susceptibility $\chi(\Vec{0},0)$ 
           against temperature for 
           the two-orbital Hubbard model on the fcc lattice
           for $U=4$ for band filling $n=1.5$(a) and $n=0.4$(b). 
           We switch on $U'$ (blue) and then $J$ (orange and red) 
           preserving the rotational symmetry ($U=U'+2J$).  
           The solid lines are guide to the eye, and
           the dashed lines extrapolations.}
  \label{fig:fccsus}	
\end{figure}

First, we discuss the results for the simple-cubic lattice (not shown).
We used $(U,U',J)=(8,6,1)$, 
which are reasonable (or somewhat overestimated) 
values for transition metals.
$\chi(\Vec{0},0)$ does not exhibit divergent behaviors at any filling
even when we extrapolate the results to low temperatures, 
which indicates that the itinerant ferromagnetism cannot be 
solely attributed to the multiorbital interactions.
This result is similar to the result for an $\infty$D
hypercubic lattice,\cite{mk98} where the ferromagnetic ground state
appears only for very large $U\gtrsim 3W$ and $J\gtrsim W/2$.

Now we turn to the fcc lattice.
Figure \ref{fig:fccsus} shows how the temperature dependence of 
$\chi(\Vec{0},0)^{-1}$ for $n=1.5, 0.4$ 
changes as we successively introduce $U'$ 
[$(U,U',J)=(4,0,0)$: single-orbital case $\rightarrow$ 
$(4,4,0)$], and then $J$ [$(4,4,0) \rightarrow 
(4,U',J)$], preserving the rotational symmetry 
($U=U'+2J$) for multiorbital cases ($U'\neq 0$).
We can see that $\chi$ is noticeably suppressed 
and the transition disappears when $U'$ alone is switched on.  
When $J$ is added, however, 
$\chi$ is significantly enhanced, 
and $T_C$ is pulled back to finite values comparable to 
that in the single-orbital case. 
The way in which $T_C$ is pulled back strongly depends on $J$: 
At $n=1.5$, $T_C\simeq 0.015$ for $J=0.25$ while 
$T_C\simeq 0.05$ for $J=0.5$.
The result clearly shows a crucial role of the interorbital 
interactions ($U'$ and $J$) in the ferromagnetism.

In Ni the triply-degenerate $t_{2g}$ bands
have 0.6 holes per site, so that 
$n=0.4$ for the present two-orbital model roughly corresponds to 
the band filling of Ni in the hole picture.  
Thus the present result, Fig.\ref{fig:fccsus}(b), 
implies a substantial role of 
Hund's coupling in the metallic ferromagnetism in the parameter 
region that contains Ni's.  This is to be 
contrasted to the Kanamori theory, to which we shall come back.

\begin{figure}[t]
  \center{\includegraphics[width=7.5cm, height=5cm]{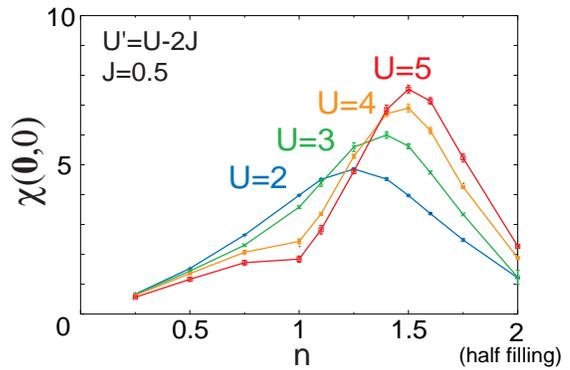}}
  \caption{(Color online) 
           Spin susceptibility against the band filling $n$ for 
           the two-orbital Hubbard model on the fcc lattice at $T=0.1$
           for various values of $U$, with
           $J=0.5$ and the relation $U'=U-2J$ fixed.}
  \label{fig:fcc_ndep}	
\end{figure}

Figure \ref{fig:fcc_ndep} displays the band-filling dependence of 
$\chi(\Vec{0},0)$ at $T=0.1$ for various values of $U = 2-5$,
where we have fixed $J=0.5$ and the relation $U'=U-2J$. 
The susceptibility is seen to 
take the largest value at around the filling 
$n_{\rm p}=1.2$-1.5, where 
$n_{\rm p}$ shifts markedly to higher $n$ as $U$ is increased.
So the filling dependence of $\chi$ is very {\it dissimilar} to 
the initial DOS (Fig.~\ref{fig:fccdos}), which has a peak 
at around $n\simeq 0.4$.  
This is to be contrasted with one-band results such as
Stoner's, $T$-matrix\cite{k63}, or FLEX theories.\cite{ao00,footnoteu98} 
The result is even distinct from multiorbital results such as 
the Gutzwiller approximation,\cite{footnotebw}
where a higher DOS at $E_F$ favors ferromagnetism. 
Hence the present result indicates that the 
correlation effect is indeed involved in the ferromagnetism.

Here we identify a component that contributes to the 
correlation effect.  
In the Kanamori theory with the $T$-matrix approximation,
which takes only particle-particle (p-p) scatterings, 
the direct interaction is reduced to $U/(1+U\chi^{\rm pp})$ while
Hund's term to 
[$J/(1+U\chi^{\rm pp})^2$], 
so the latter is concluded to be less relevant. 
However, while the $T$-matrix approximation is justified only 
for low electron densities, other types of scatterings 
such as the particle-hole (p-h) channel\cite{ao00,l81} can play a role
for general band fillings.
In particular, the p-h scatterings enhance $U$ into 
$U/(1-U\chi^{\rm ph})$ ($\chi^{\rm ph}$: the p-h propagator), 
and $J$ is even more enhanced into  [$J/(1-U\chi^{\rm ph})^2$].  
Since the present DMFT calculation effectively 
includes these p-h scatterings, this should be one 
component that substantially contributes to the ferromagnetism.

\begin{figure}[t]
  \center{\includegraphics[width=8cm, height=5cm]{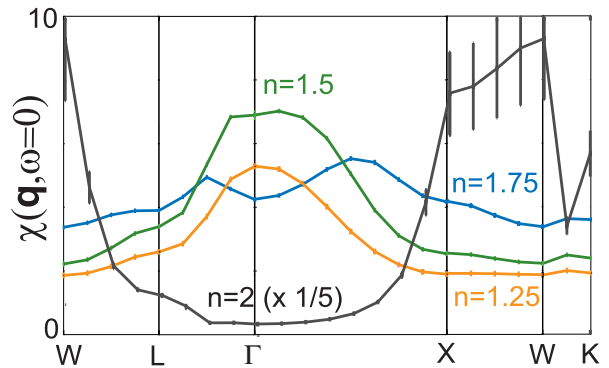}}
  \caption{(Color online) 
  Wave number dependence of the susceptibility on the fcc lattice 
  for various values of 
  $n=1.25 - 2.0$ for $U=4,~U'=3,~J=0.5,~T=0.1$.}
  \label{fig:susc_q}
\end{figure}

The interaction $\hat{H}_{\rm int}$ should become more effective,
especially in the p-h channel,
as the band filling approaches the half filling ($n=2$). 
This will enhance $\chi$, as actually seen in Fig.~\ref{fig:fcc_ndep}
for $n<n_{\rm p}$.
$n_{\rm p}$ shifts to higher densities with $U$.  
This should be because 
the enhancement of $\chi$ with $U$ becomes stronger toward the 
half filling, while the original DOS is conversely 
peaked around low $n$. 
On the other hand, $\chi$ starts to 
decrease for $n>n_{\rm p}$, which is expected 
to come from antiferromagnetic correlations.
To confirm the antiferromagnetic correlations,
we have calculated the wave number dependence of 
$\chi(\Vec{q},0)$, where the $\Vec{q}$ dependence 
incorporated through $\hat{\c}_0$ in Eq.~(\ref{eq:bs}). 
We can see that $\chi$ in Fig.~\ref{fig:susc_q}
has a peak at around ${\rm \Gamma}$ point 
for $n=1.25$ and 1.5, while the peak disappears for $n=1.75$ 
in favor of a diverging peak at around X and W points 
for $n=2$, which indicates 
strong antiferromagnetic correlations as the half filling is approached.

\begin{figure}[t]
  \center{\includegraphics[width=8.2cm, height=4cm]{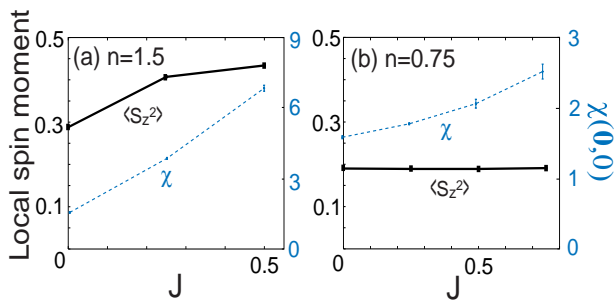}}
  \caption{(Color online) Local spin moment (black) versus $J$ in 
           the two-orbital Hubbard model on the fcc lattice
           for (a) $n=1.5$ and (b) $n=0.75$ with $T=0.1, U=4$
           and the relation $U'=U-2J$ fixed.
           The corresponding susceptibility $\c(\Vec{0},0)$ 
           is plotted in blue.
           }
  \label{fig:mom}	
\end{figure}

We make another argument from the the strong-coupling limit,
where the Hamiltonian is given only by the intra-atomic interaction
$\hat{H}_{\rm int}$.
The two-electron eigenstates are then classified
by the spin and orbital symmetries as

\vspace{5pt}
\begin{tabular}{ccccc}
  Notation & Spin    & Orbital & Expression & Energy \\ \hline
  1S$^\pm$ & singlet & sym. & $\frac{1}{\sqrt{2}}
  (c_{1\ua}^\dagger c_{1\da}^\dagger\pm c_{2\ua}^\dagger c_{2\da}^\dagger)
  $ & $U\pm J$\\
  1S$^0$ & singlet & sym. & $\frac{1}{\sqrt{2}}
  (c_{1\ua}^\dagger c_{2\da}^\dagger+c_{2\ua}^\dagger c_{1\da}^\dagger)
  $ & $U'+J$\\
  3$^0$A & triplet & antisym. & $\frac{1}{\sqrt{2}}
  (c_{1\ua}^\dagger c_{2\da}^\dagger-c_{2\ua}^\dagger c_{1\da}^\dagger)
$ & $U'-J$\\
  3$^\s$A & triplet & antisym. & $c_{1\s}^\dagger c_{2\s}^\dagger
$ & $U'-J$.\\
\end{tabular}
\vspace{3pt}

\noindent While all the spin-triplet states have the same energy $U'-J$,
the orbital-symmetric states split into three energies, 
$U+J$, $U'+J$ and $U-J$.
Then the ground state for $(U,U',J)=(4,0,0)$ is
a superposition of the four states, 1S$^0$, 
and 3$^{0,\ua,\da}$A,
while that for $(4,4,0)$ becomes 
a superposition of the six states, 1S$^{0,\pm}$ and 3$^{0,\ua,\da}$A.
Namely, the latter has three times greater number of 
spin singlets than the former.
This intuitively accounts for the reduction of $\c(\Vec{0},0)$ 
with $U'$ in Fig.~\ref{fig:fccsus}.  

In the strong-coupling limit,
$J$ further brings down the energy of the spin triplets.
This may be one reason for the increase of $\c(\Vec{0},0)$
with $J$ in Fig.~\ref{fig:fccsus}:
The local spin moment, $\langle S_{z}^2\rangle$, in Fig.~\ref{fig:mom}
actually increases with $J$ for $n=1.5$.
For $n=0.75$, however, the local moment remains almost constant while
$\chi(\Vec{0},0)$ noticeably increases with $J$.
This implies that Hund's coupling, despite being a local interaction, 
aligns spins nonlocally via the one-electron hoppings.

Future problems include the effect of 
anisotropic transfers or hybridization 
of $d$ orbits, and an examination of the chemical trend in the itinerant 
ferromagnetism.

We thank Masatoshi Imada and Kazuma Nakamura for valuable discussions.  
This work is in part supported by a Grant-in-Aid for Science Research on
Priority Area ``Anomalous quantum materials" 
from the Japanese Ministry of Education.
The calculations are partly done at the Supercomputer Center,
ISSP, University of Tokyo.


\begin{references}
\bibitem{footnoteu98} 
Although the present filling dependence happens to be similar to a 
single-orbital DMFT result\cite{u98}, 
the multiorbital situation is 
dominated by a balance between $U'$ and $J$ as emphasized here.
\end{references}
\end{document}